\documentclass[manuscript,screen]{acmart}

\DeclareRobustCommand{\bbone}{\text{\usefont{U}{bbold}{m}{n}1}}

\usepackage{algorithm} 
\usepackage{algpseudocode}

\usepackage{dsfont}

\usepackage{caption}
\usepackage{subcaption}
\usepackage{multirow}
\usepackage{graphicx,wrapfig,lipsum}
\usepackage{float}
\usepackage[absolute]{textpos}
\usepackage{bbm}
\usepackage[english]{babel}
\usepackage{wrapfig}
\usepackage{bm}

\AtBeginDocument{%
  }

\setcopyright{acmcopyright}
\copyrightyear{2023}
\acmYear{2023}
\acmDOI{XXXXXXX.XXXXXXX}

\acmConference[CONSEQUENCES@RecSys '23]{ACM Conference on Recommender Systems}{September 18--22,
  2023}{Singapore}

\acmPrice{15.00}
\acmISBN{978-1-4503-XXXX-X/18/06}

\setcopyright{none}
\renewcommand\footnotetextcopyrightpermission[1]{}
\begin{document}

\begin{textblock}{10}(3,0.4)
 \noindent\normalsize  \begin{center}CONSEQUENCES Workshop on Causality, Counterfactuals, and Sequential Decision-Making \\ in conjunction with ACM RecSys 2023\end{center}
 \end{textblock}

\title{Fairness of Exposure in Dynamic Recommendation}

\author{Masoud Mansoury}
\affiliation{%
  \institution{University of Amsterdam,}
  \institution{Elsevier Discovery Lab}
  \city{Amsterdam}
  \country{Netherlands}}
\email{m.mansoury@uva.nl}

\author{Bamshad Mobasher}
\affiliation{%
  \institution{DePaul University}
  \city{Chicago}
  \country{USA}}
\email{mobasher@cs.depaul.edu}


\renewcommand{\shortauthors}{Mansoury and Mobasher}

\begin{abstract}
  Exposure bias is a well-known issue in recommender systems where the exposure is not fairly distributed among items in the recommendation results. This is especially problematic when bias is amplified over time as a few items (e.g., popular ones) are repeatedly over-represented in recommendation lists and  users' interactions with those items will amplify bias towards those items over time resulting in a \textit{feedback loop}. This issue has been extensively studied in the literature in static recommendation environment where a single round of recommendation result is processed to improve the exposure fairness. However, less work has been done on addressing exposure bias in a dynamic recommendation setting where the system is operating over time, the recommendation model and the input data are dynamically updated with ongoing user feedback on recommended items at each round. In this paper, we study exposure bias in a dynamic recommendation setting. Our goal is to show that existing bias mitigation methods that are designed to operate in a static recommendation setting are unable to satisfy fairness of exposure for items in long run. In particular, we empirically study one of these methods, \textit{Discrepancy Minimization}, and show that repeatedly applying this method fails to fairly distribute exposure among items in long run. To address this limitation, we show how this method can be adapted to effectively operate in a dynamic recommendation setting and achieve exposure fairness for items in long run. Experiments on a real-world dataset confirm that our solution is superior in achieving long-term exposure fairness for the items while maintaining the recommendation accuracy.
\end{abstract}




\maketitle

\section{Introduction}

Recommender systems are known to suffer from \textit{exposure bias}; few items are over-exposed in the recommendation lists, while the majority of other items are under-exposed \cite{mansoury2022understanding,abdollahpouri2020multi,wu2022joint,khenissi2020modeling}. 
A negative consequence of this bias is that it may lead to unfairness on supplier-side: items belonging to different suppliers are unfairly represented in the recommendations results. Also, due to the \textit{feedback loop} phenomena in which users profiles get updated over time via potentially biased recommendations, exposure bias can be amplified over time \cite{mansoury2020feedback,sinha2016deconvolving}. Highly exposed items in the recommendation lists have higher chance to be viewed/examined/clicked by the users, while insufficiently exposed items will not receive adequate attention from the users. As a result, those over-exposed items would have a higher chance to be shown in the recommendation lists in the future which amplifies existing bias. Amplification of exposure for few items would be at the expense of the under-exposure for a majority of other items (the ones that might be interesting for some users).

Most existing research to study exposure bias are conducted in static recommendation setting where a single round of recommendation results is analyzed \cite{patro2020fairrec,mansoury2021graph,mansoury2021fairness,mansoury2022understanding,suhr2019two}. While these studies unveiled important aspects of exposure bias and proposed solutions for tackling it, the long-term impact of this bias is yet a significant research gap which we seek to investigate in this paper. Filling this gap requires studying the task of recommendation problem in dynamic setting where the recommendation process is repeatedly performed over time \cite{mansoury2022exposure,mansoury2021unbiased,li2010contextual,li2020cascading,hiranandani2020cascading,mansoury2020feedback}. At each round $t$, the recommendation lists are generated for all users, clicks on recommended items are collected, and users' profile are updated. Updated users' profile are used for recommendation process in round $t+1$. 

In this paper, we empirically study exposure bias in dynamic recommendation setting. First, we show that existing bias mitigation algorithms designed to operate in a static recommendation setting fail to satisfy exposure fairness in long run, when used to generate recommendations for the users over time. This is done by simulating the recommendation process with bias mitigation algorithm in \cite{antikacioglu2017post}. The experimental results suggest that a different treatment is needed for improving the exposure fairness in long run. Therefore, we modify the algorithm to adapt it for improving exposure fairness in dynamic recommendation environment. To do so, instead of solely distributing the available exposure in each round among items, we modify the algorithm to distribute the available exposure in each round $t$ proportional to the cumulative exposure given to the items up to round $t$. This will ensure to avoid over-exposure or under-exposure in long run. Experiments on a publicly-available dataset confirm that the modified algorithm significantly outperforms the original algorithm in terms of exposure fairness in long run, while maintaining the accuracy of the recommendations.      


\begin{figure}
    \begin{subfigure}[b]{0.8\textwidth}
        \includegraphics[width=\textwidth]{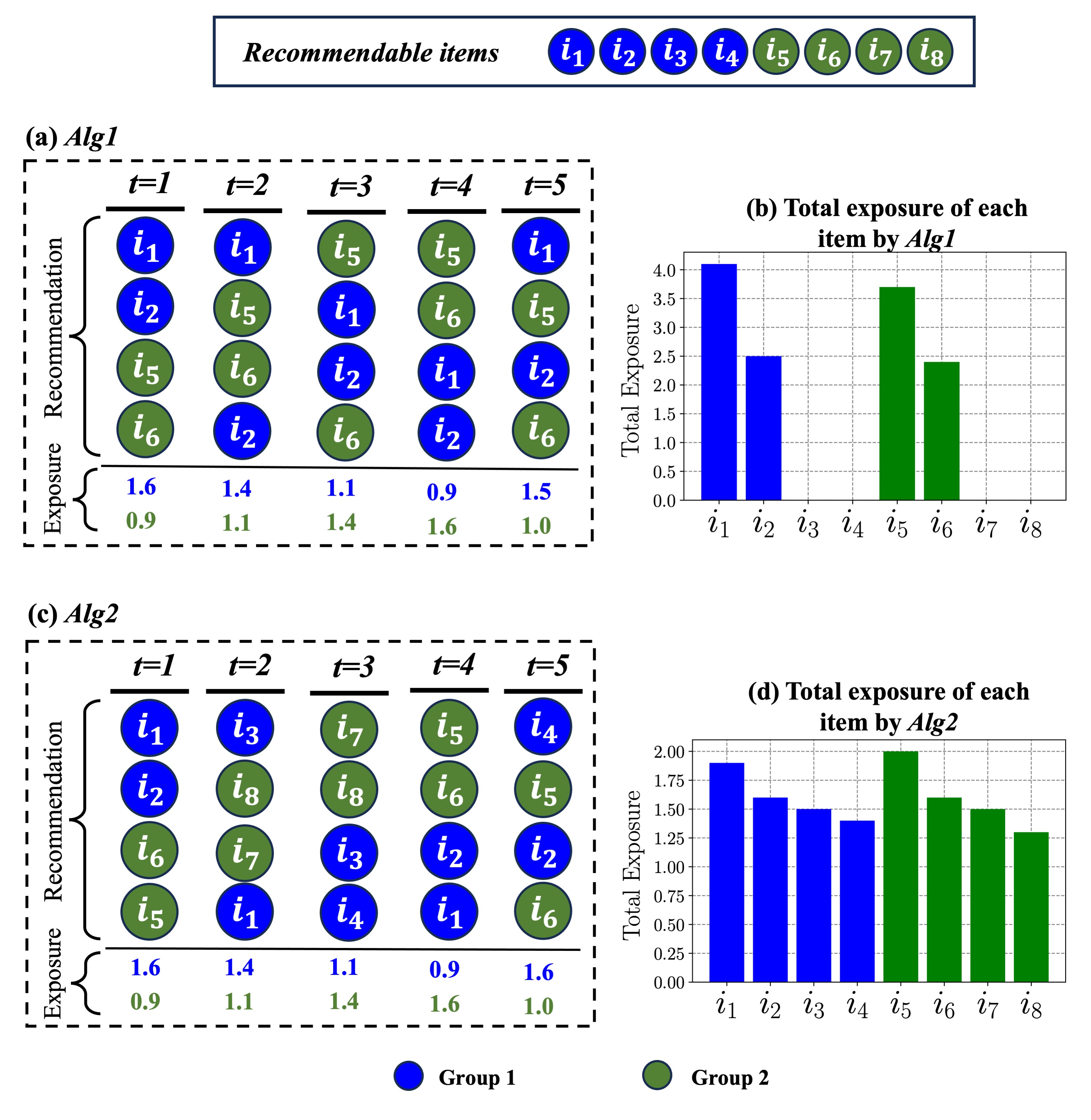}
    \end{subfigure}
\caption{Comparison between bias mitigation algorithms designed for static recommendation setting (\textit{Alg1}) and dynamic recommendation setting (\textit{Alg2}).}\label{fig_example}
\end{figure}

\section{Motivating example}

Figure \ref{fig_example} illustrates a situation where tackling exposure bias in static setting, while achieves fairness objectives in each single round of the recommendation results, fails to ensure fairness in long run. Given items in candidate pool, the goal is
to extract a list of items (in this example, size of 4) as recommendation to the user that satisfies exposure fairness among the items. Hence, items are categorized into two groups in which one group can be interpreted as unprotected items that are often over-exposed in the recommendation results (e.g., popular items or items that represent the majority), and another group that can be interpreted as protected items that are often under-exposed in the recommendation results (e.g., unpopular items or items that represent the minority). The recommendation process is repeated for 5 rounds. We assume that the results are obtained from two different bias mitigation algorithms: \textit{Alg1} attempting to improve the exposure fairness in static setting and \textit{Alg2} attempting to improve exposure fairness in dynamic setting.

Figure \ref{fig_example}a shows the results obtained by \textit{Alg1}. As shown, in each round, items belonging to each group are almost fairly represented: 50\% of the recommendation slots are assigned to \textit{Group 1} and 50\% to \textit{Group 2}. In terms of the position in the recommendation list, items from both groups are almost fairly represented with \textit{Group 1} being exposed slightly more in round 1 and \textit{Group 2} being exposed slightly more in round 4. This is the way that existing bias mitigation approaches attempt to improve the exposure fairness in recommendation results. While they properly achieve exposure fairness in each single round of recommendations, their performance in long run is still unclear. Figure \ref{fig_example}b shows the total exposure assigned to each item in long run. As shown, most of the exposure are accumulated on few items ($i_1, i_2, i_5,i_6$). This is mainly due to the fact that the algorithm is not aware of the history of the exposure for each item in the past rounds and only attempts to distribute the exposure of a single round of the recommendation among items.

In contrast to \textit{Alg1}, \textit{Alg2} operates in a dynamic recommendation setting. Figure \ref{fig_example}c shows the results obtained by \textit{Alg2}. 
Analogous to \textit{Alg1} in Figure \ref{fig_example}a, it yields the same exposure fairness for each group of items in each round. However, unlike \textit{Alg1}, it provides almost equal exposure to each item in long run. Figure \ref{fig_example}d shows the total exposure given to each item by \textit{Alg2}. It is evident that in long run items are more equally represented in the recommendation results.

\section{Formalizing Item Exposure in Dynamic Recommendation Setting}\label{exposure}

Given $\mathcal{U}=\{u_1,...,u_n\}$ as the set of $n$ users, $\mathcal{I}=\{i_1,...,i_m\}$ as the set of $m$ items, the recommendation task at each round $t=\{1,...,T\}$ is retrieving a ranked list of $K$ most relevant items for each user. We denote the recommendation lists generated at round $t$ as $R^t$, with $R^t_u$ being the list generated for user $u$ at round $t$.

Exposure of an item in the recommendation results depends on its position in the list \cite{singh2018fairness}. 
It is possible an item gets recommended, but receives no exposure because it was probably shown at the bottom of the list and the user has not examined it. For this reason, items appearing on top of the list receive higher exposure than items appearing at the bottom of the list. 
Following the idea in \cite{singh2018fairness}, we consider a standard exposure drop-off to assign weight to each position in the recommendation lists as commonly used in ranking metrics (e.g., nDCG) and calculated as $1/\log(1+k)$ where $k$ is the position of the item in the list. Therefore, we compute the exposure of an item $i$ in round $t$ as:

\begin{equation}\label{eq_e}
    E^t(i) = \sum_{u \in \mathcal{U}}\sum_{k=1}^{K}\mathds{1}(i = R^t_{u}(k)).\frac{1}{\log_2(1+k)}
\end{equation}

\noindent where $R^t_{u}(k)$ is the $k$-th item in $R^t_u$ and $\mathds{1}(.)$ is the indicator function returning zero when its argument is False and 1 otherwise. We define the cumulative exposure of an item in $T$ rounds of the recommendation process as:

\begin{equation}\label{eq_te}
    E^{\leq T}(i) = \sum_{t=1}^T{E^t(i)}
\end{equation}

\section{Mitigating Exposure Bias in Dynamic Recommendation Setting}\label{sec_dm}

In this section, we review an existing bias mitigation method, \textit{Discrepancy Minimization (DM)}, proposed in \cite{antikacioglu2017post} and then show how it can be adapted to achieve exposure fairness in long run. \textit{DM} is a post-processing approach that operates on a bipartite graph constructed over the long recommendation lists. This bipartite recommendation graph with users and items as its nodes and edges between them representing the recommendation is further processed to construct the flow network. Then, the \textit{minimum-cost network flow} methods \cite{ahuja1988network} are applied on the constructed flow network to find the recommendation subgraphs that yields exposure fairness for items. More specifically, an integer-valued constraint is defined as the target exposure for each item (i.e., the exposure that we desire for each item), and the minimum-cost flow network is optimized to find an optimal subgraph that gives the minimum discrepancy from this constraint. 

For simplicity, we assume \textit{Equality of Exposure} among items as the fairness notion in which the goal is to \textit{equally} distribute exposure among items. Therefore, the constraint is defined as:

\begin{equation}\label{eq_c}
    \mathcal{C}^t(i) = \left\lfloor\frac{n \times K}{\sum_{i \in \mathcal{I}}{\bbone(i \in \mathcal{L})}}\right\rfloor
\end{equation}

\noindent where $\mathcal{L}$ is the long recommendation lists generated by a \textit{base recommender} for all users. In this equation, the nominator is the total number of available recommendation slots and the denominator is the total number of items in the long recommendation lists. Hence, Eq. \ref{eq_c} equally assigns the recommendation slots to the items. Depending on the relevancy of the items, this constraints may not be perfectly satisfied, but \textit{DM} attempts to minimize the discrepancy with this constraint while maintaining the recommendation accuracy. 

The constraint in Eq. \ref{eq_c} is designed for static recommendation setting which separately distributes the available recommendation slots in each single round of recommendation process. Although it shows promising results in mitigating exposure bias, we show that it can be even further improved by adapting it to operate in dynamic recommendation setting. Therefore, we define the constraint for dynamic recommendation setting as:

\begin{equation}\label{eq_dc}
    \mathcal{C}^{\leq t}(i) = \left\lfloor\frac{\mathcal{C}^t(i)}{E^{\leq t}(i)}\right\rfloor
\end{equation}

\noindent This way, the exposure given to each item at round $t$ is proportional to the total exposure given to that item up to round $t$. This means that if an item is in total over-exposed up to round $t$, Eq. \ref{eq_dc} assigns less exposure to it in round $t$, and vice versa. For the rest of the paper, we denote $DM^{static}$ and $DM^{dynamic}$ for $DM$ with constraint in Eq. \ref{eq_c} and \ref{eq_dc}, respectively.

\section{Experiments}

We follow the simulation process in \cite{mansoury2020feedback} for our experiment. In each round $t$, the users' profile in input data is randomly divided into 80\% as training set and 20\% as test set. The training set is used to generate the recommendation lists (first long recommendation lists by a base recommender and then final recommendations by a reranker) and the test set is used to evaluate them. Acceptance probability is used to collect users' clicks on recommended items as $prob(i|R_u^t)=e^{\alpha \times rank_i}$ where $\alpha$ is a negative value ($\alpha < 0$) for controlling the probability assigned to each recommended item and $rank_i$ is the position of item $i$ in $R_u^t$. We set $\alpha=-0.5$ for our experiments. Collected clicks are added to the users' profile and the updated input data is used for round $t+1$. This process repeats for $T$ rounds.

We perform our experiments on MovieLens 1M dataset \cite{harper2015movielens}. In each round, we use Matrix Factorization (\textit{MF}) \cite{koren2009matrix} for generating the long recommendation lists of size 40 for each user. Then, we apply the bias mitigation methods described in section \ref{sec_dm} to generate the final recommendation lists of size 10 for each user. The hyperparameters involved in both \textit{MF} and \textit{DM} are tuned for the highest possible accuracy. We use Normalized Discounted Cumulative Gain (nDCG) as the accuracy metric to measure the ranking quality of the recommendation. We perform the simulation using \textit{librec-auto} recommendation tool \cite{mansoury2019algorithm} for $T=600$ rounds.


\begin{figure}
    \includegraphics[width=0.7\textwidth]{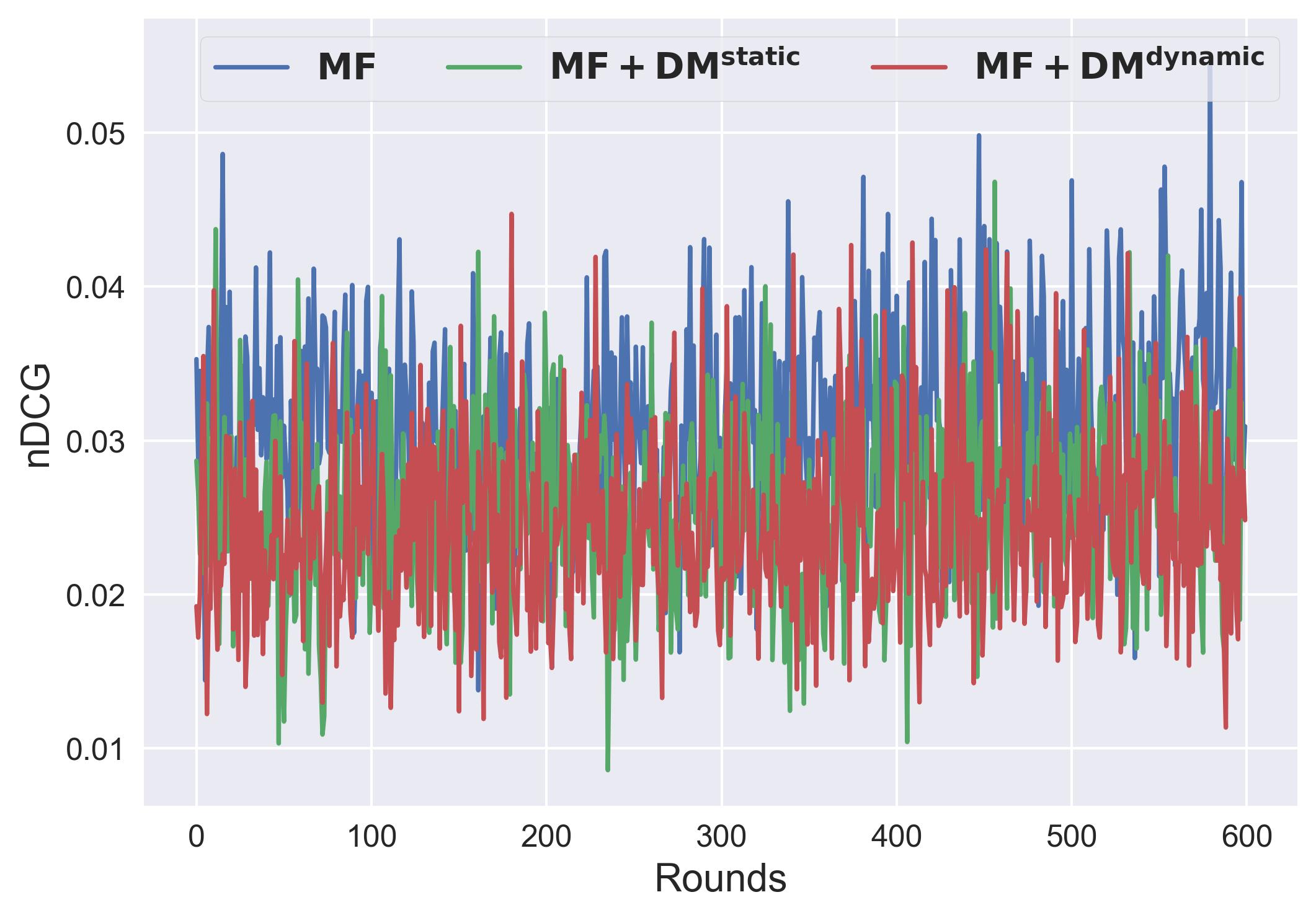}
\caption{Ranking quality of the recommendation results.}\label{fig_ml_results_a}
\end{figure}

\subsection{Experimental results}

In this section, we present our experimental results. We compare the performance of three models: 1) \textit{MF} recommendation model ($\mathbf{MF}$), 2) \textit{MF} as a base recommender to generate the long recommendation lists and $DM^{static}$ as abias mitigation method to generate the final recommendation lists ($\mathbf{MF + DM^{static}}$), 3) similarly, \textit{MF} as a base recommender and $DM^{dynamic}$ as a bias mitigation method ($\mathbf{MF + DM^{dynamic}}$). Figure \ref{fig_ml_results_a} compares the models in terms of nDCG in each round. As shown, \textit{MF} yielded slightly higher nDCG than both \textit{DM} variations, indicative of insignificant loss in ranking quality when applying these bias mitigation methods.

\begin{figure}
    \begin{subfigure}[b]{0.45\textwidth}
        \includegraphics[width=\textwidth]{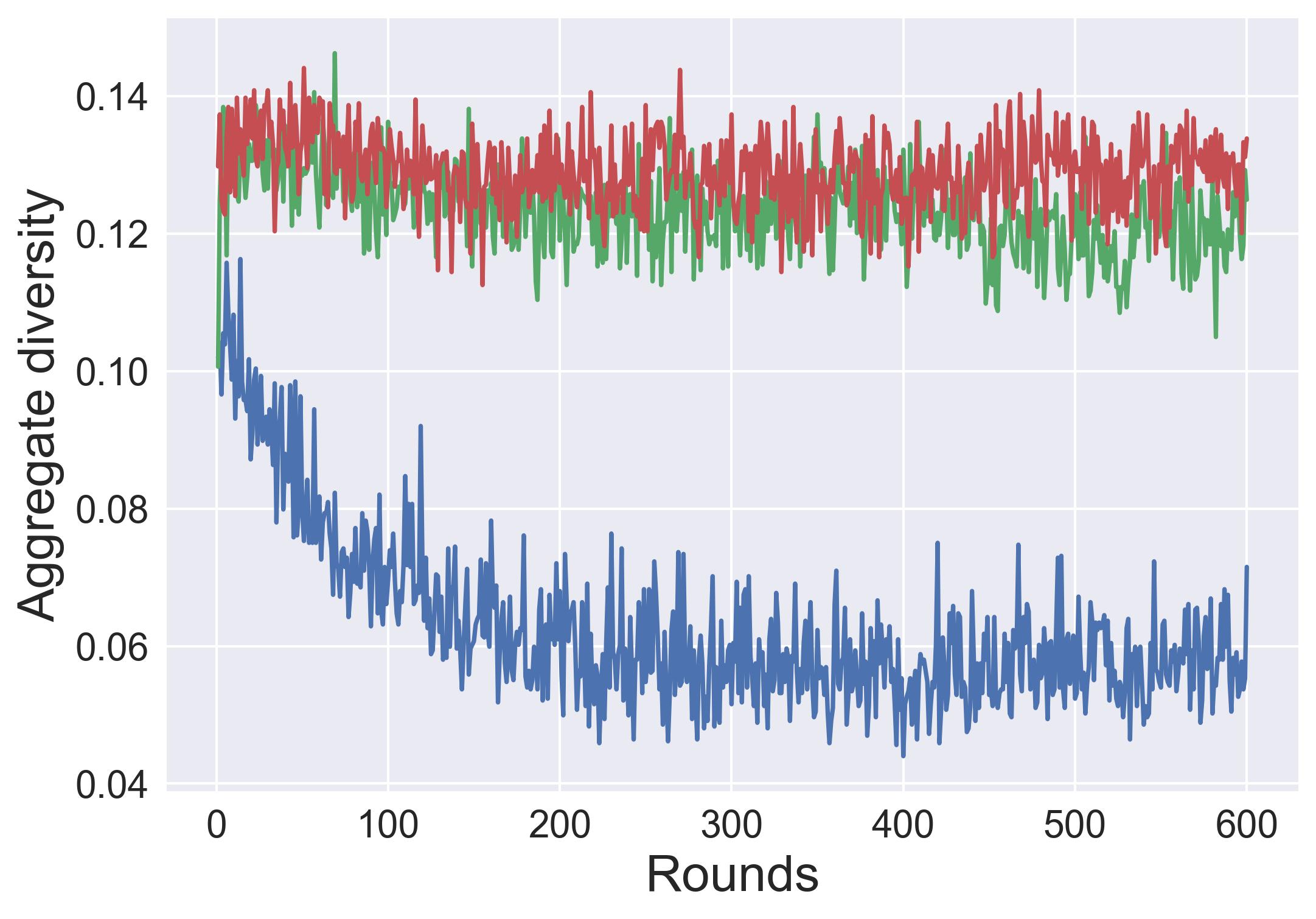}
        \caption{Aggregate diversity per round}\label{fig_ml_results_aggdiv_a}
    \end{subfigure}
    \begin{subfigure}[b]{0.45\textwidth}
        \includegraphics[width=\textwidth]{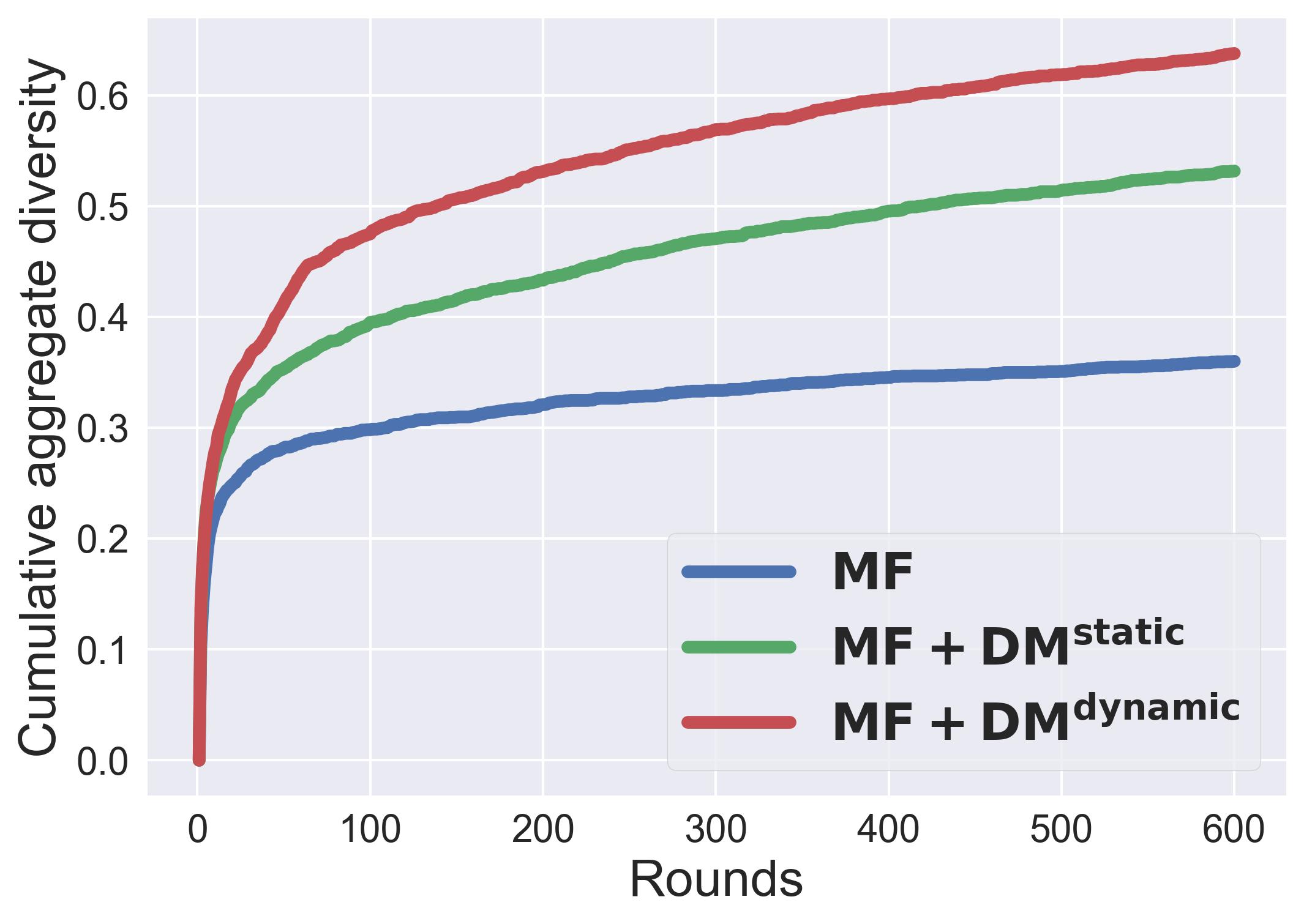}
        \caption{Cumulative Aggregate diversity}\label{fig_ml_results_aggdiv_b}
    \end{subfigure}
\caption{Experimental results on MovieLens dataset on the performance of recommendations in terms of \textit{aggregate diversity} using pure \textit{MF} algorithm and \textit{MF} as base recommender and $DM^{static}$ and $DM^{dynamic}$ as post-processing methods.}\label{fig_ml_results_aggdiv}
\end{figure}

Figure \ref{fig_ml_results_aggdiv} shows the results in terms of \textit{aggregate diversity} \cite{adomavicius2011improving}: the percentage of items which appear at least once in the recommendation lists. Figure \ref{fig_ml_results_aggdiv_a} shows the aggregate diversity of the recommendations per round. The results suggest that both $DM$ variations significantly outperformed \textit{MF} and also our $DM^{dynamic}$ slightly achieved fairer exposure outputs compared to $DM^{static}$. Our main results are shown in Figure \ref{fig_ml_results_aggdiv_b} which presents the cumulative aggregate diversity of the recommendations in long run: it shows the percentage of the items appeared at least once in $t$ rounds of the recommendation process. As shown, our $DM^{dynamic}$ outperformed both \textit{MF} and $DM^{static}$ in covering more items in the recommendation lists in long run.

\begin{figure}
    \begin{subfigure}[b]{0.45\textwidth}
        \includegraphics[width=\textwidth]{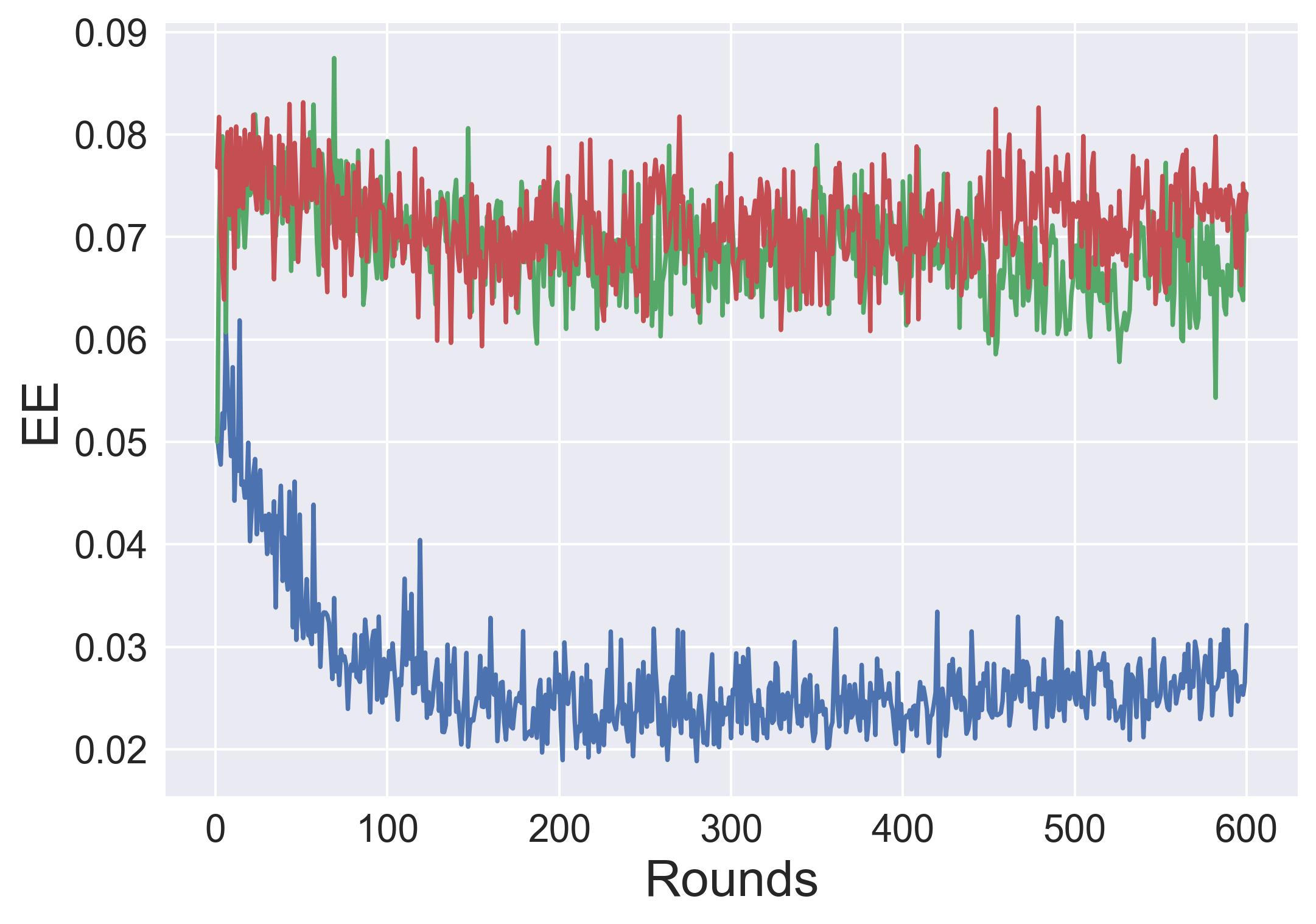}
        \caption{Equality of Exposure}\label{fig_ml_results_ee_a}
    \end{subfigure}
    \begin{subfigure}[b]{0.45\textwidth}
        \includegraphics[width=\textwidth]{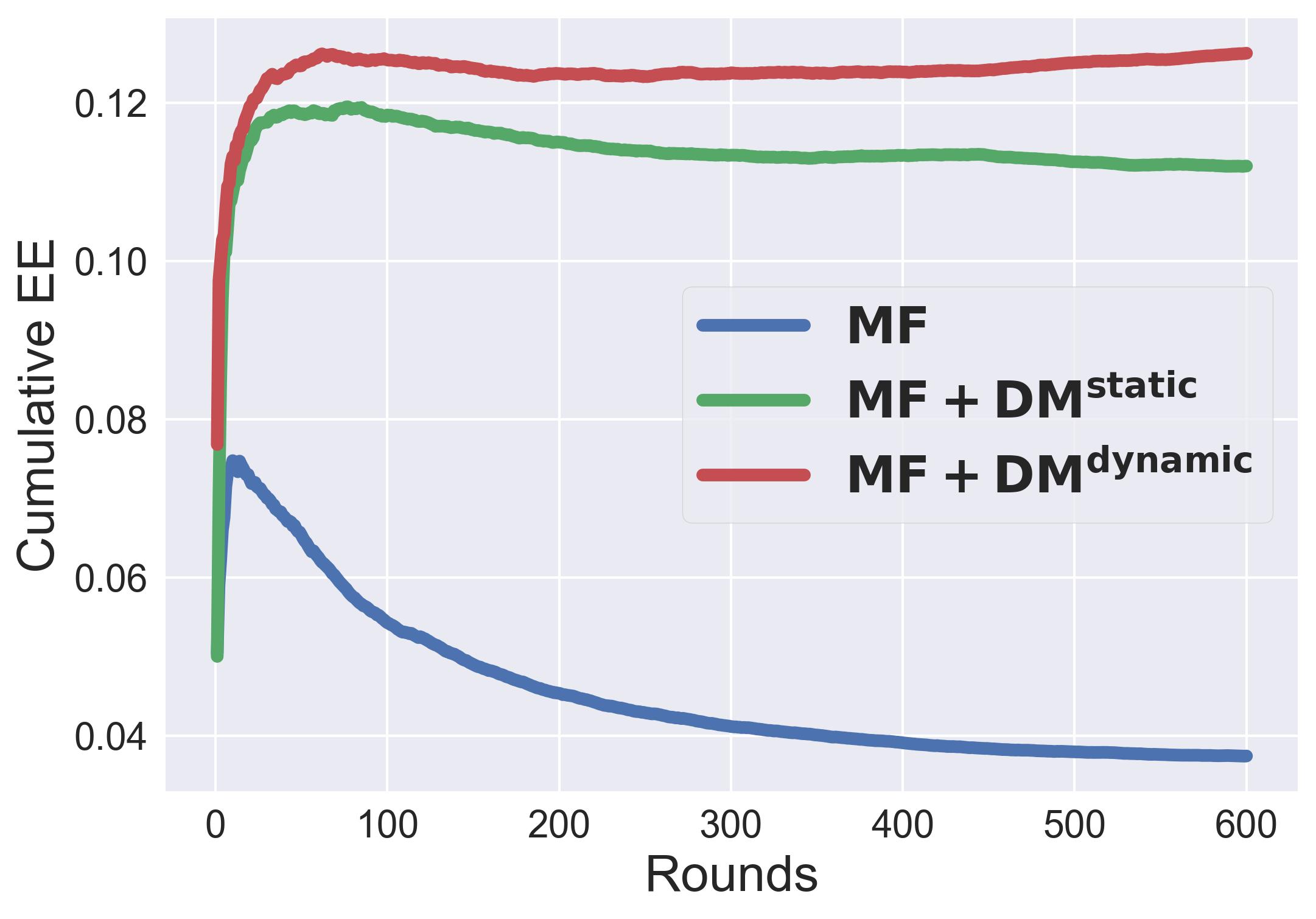}
        \caption{Cumulative EE}\label{fig_ml_results_ee_b}
    \end{subfigure}
\caption{Experimental results on MovieLens dataset on the performance of recommendations in terms of \textit{equality of exposure} using pure \textit{MF} algorithm and \textit{MF} as base recommender and $DM^{static}$ and $DM^{dynamic}$ as post-processing methods.}\label{fig_ml_results_ee}
\end{figure}

Figure \ref{fig_ml_results_ee} shows the results in terms of \textit{Equality of Exposure} \cite{mansoury2022understanding}: how equal items are represented in recommendation results. Given the distribution of items exposure in round $t$ computed using Eq. \ref{eq_e}, we use \textit{Gini Index} \cite{vargas2014improving} to measure the uniformity of this distribution. Uniform distribution will have Gini Index equal to zero which is the ideal case. Therefore, for consistency, we report $1 - Gini Index$ as equality of exposure. Figure \ref{fig_ml_results_ee_a} shows the equality of exposure for each recommendation results per round. As shown, our $DM^{dynamic}$ slightly resulted in fairer exposure for items in each round. In terms of long-term performance, Figure \ref{fig_ml_results_ee_b} presents our main results, the equality of exposure cumulatively in long run. Here, the Gini Index is computed over the distribution of items exposure derived by Eq. \ref{eq_te}. Clearly, our $DM^{dynamic}$ is superior in improving exposure fairness in long run compared to other methods as it yielded higher EE compared to other recommendation methods.

\section{Conclusion}

In this paper, we studied exposure bias in dynamic recommendation setting through an offline simulation. We hypothesized that bias mitigation methods designed to operate in a static recommendation setting are unable to sufficiently address exposure bias in long run. We based our analysis on one of these bias mitigation methods, \textit{Discrepancy Minimization}, and empirically showed the validity of our claim on this algorithm. To tackle this limitation, we proposed solution for adapting this algorithm to operate in dynamic recommendation environment. Experiments on a movie recommendation dataset showed the superiority of our solution in improving exposure fairness in long run. 


\bibliographystyle{ACM-Reference-Format}
\bibliography{ref}

\end{document}